\documentclass[twocolumn,prb,amsmath,amssymb]{revtex4}

\usepackage{graphicx}
\usepackage{dcolumn}
\usepackage{bm}

\begin{document}


\title{On the properties of superconducting planar resonators at mK temperatures}

\author{T. Lindstr\"{o}m}
\email{tobias.lindstrom@npl.co.uk}
\affiliation{University of Birmingham, Edgbaston, Birmingham B15 2TT, UK}
\altaffiliation [Also at ]{National Physical Laboratory, Hampton Road, Teddington, TW11 0LW, UK}

\author{J.E. Healey}
\affiliation{University of Birmingham, Edgbaston, Birmingham B15 2TT, UK}%

\author{M.S. Colclough}
\affiliation{University of Birmingham, Edgbaston, Birmingham B15 2TT, UK}%

\author{C.M. Muirhead}
\affiliation{University of Birmingham, Edgbaston, Birmingham B15 2TT, UK}%

\author{A.Ya.Tzalenchuk}

\affiliation{National Physical Laboratory, Hampton Road, Teddington, TW11 0LW, UK}

\date{\today}

\begin{abstract}
Planar superconducting resonators are now being increasingly used
at mK temperatures in a number of  novel applications. They are
also interesting devices in their own right since they allow us to
probe the properties of both the superconductor and its
environment. We have experimentally investigated three types of
niobium resonators - including a lumped element design -  fabricated
on sapphire and SiO$_2$/Si substrates.  They all exhibit a
non-trivial temperature dependence of their centre frequency and
quality factor. Our results shed new light on the interaction
between the electromagnetic waves in the resonator and two-level
fluctuators in the substrate.
\end{abstract}

\pacs{Valid PACS appear here}
\maketitle
\section{Introduction}
The superconducting microwave resonator is an ubiquitous device
with uses ranging from the very practical - such as filters for
telecommunications - to rather exotic such as tests of cavity
quantum electrodynamics. Over recent years there has been a
resurgence in the interest in on-chip resonators; the primary
reason being their use as very sensitive photon and much of the work to date has been motivated by
the need to understand noise properties of kinetic inductance
detectors (KID)\cite{Gao,Kumar,Barends}. Resonators are also
 being used as elements in  circuitry for quantum
information processing\cite{wallraff}. The latter has
triggered a wave of investigations into the properties of
resonators when operated at mK temperatures and
very low (ideally single photon) microwave powers. Of particular
interest has been the effects of two-level fluctuators (see e.g.
\cite{clarke,faoro} and references therein).
Superconducting resonators at microwave frequencies differ from
 their normal metal counterparts primarily because of their high quality factor (Q) and a relatively large
kinetic inductance. For example, for an ideal high-Q LC
resonator we can write the centre frequency as
$f_0=(2\pi\sqrt{(L+L_K)C})^{-1}$ where $L_K$ is the contribution
from the kinetic inductance. $L_K$  depends on the order parameter and therefore varies with temperature, but can also be altered e.g. by a weak
magnetic field\cite{Healey}. As the temperature goes down the
 centre frequency increases monotonically while the conductor losses decrease
(leading to a higher quality factor), as described by the Mattis-Bardeen theory\cite{MB}. However,
at temperatures $\sim T_c/10$, these mechanisms
saturate (in Nb and its compounds around 1 K) and other effects
become prominent \cite{Barends,Gao,Healey}, most notably a
\textit{decrease} in the centre frequency as the temperature is
reduced. This 'back-bending' is now widely believed to be due to
the presence of two-level fluctuators (TLF) in the substrate.

Here we will discuss measurements of the centre frequency and
losses in three types of Nb resonators: conventional $\lambda/2$
and $\lambda/4$ geometric resonators and a recently developed type
of lumped element resonator\cite{doyle}.

\section{Theory}

A generic expression for the transmittance, $S_{21}$, of a
transmission line shunted by a resonator (e.g. the lumped element
and $\lambda/4$ resonators considered here) can be written
$S_{21}=2\left[2+\frac{g}{1+2jQ_u\delta f}\right]^{-1}$
where $Q_u$ is the unloaded quality factor ($Q$ in the absence of
coupling to the feedline), $g$ is a coupling parameter $\propto Q_u$ - the exact
expression for which will depend on the type of resonator used-
and $\delta f=(f-f_0)/f_0$ is the fractional shift from resonance.
The value of $Q_u$ is in general given by summing over all loss
mechanisms $\sum_{k} Q_k^{-1}$ but losses in the dielectric and
the superconducting film itself\cite{plourde} usually dominate in
planar resonators and radiation losses etc. can be neglected.
Microwave losses are usually quantified by  the loss tangent $\tan\delta=1/Q_u$, which in a dielectric is
proportional to the sum of the various absorbtion mechanisms $\sum_k \alpha_k$.

At sub-Kelvin temperatures one would expect the properties of
a dielectric to stay essentially constant. However, experimentally
one finds a variation in both the effective dielectic constant and
losses.  Much of this can be explained if one assumes the presence
of a bath of TLF that couple to the resonator via their electric
dipole moment. Many of the details of the corresponding theory were
worked out over 30 years ago when the effects of TLF in glasses at
low temperatures were being investigated\cite{Phillips}. The theory is
based on the application of the Bloch equations to an ensemble of
spin systems and while it is simple it nevertheless captures most of the essential physics.
For resonant absorption due to an ensemble of TLF  $\alpha_r$ has the form
 \cite{strom,schickfus}
\begin{equation}
\alpha_r=\frac{\pi \omega n d^2}{3c_0 \epsilon_0\epsilon_r}
\left(1+\frac{P}{P_c}\right)^{-1/2} \tanh \left(\frac{\hbar
\omega}{2k_BT} \right) \label{eq:absorption}
\end{equation}
where $\omega$ is the measurement frequency, $d$ is the dipole
moment, $c_0$ is the speed of light in vacuum and $n$ is the density
of states of the TLF that couple to the stray field of the
resonator. Note that this density should in principle depend on
frequency, but is experimentally often found to be constant.
The temperature dependence of the
absorption reflects the population difference of the two states of
the fluctuators, as the temperature increases the TLF will therefore absorb \textit{less} power.
Note that Eq.~1 only accounts for \textit{resonant} absorbtion,  non-resonant relaxation processes due the TLF bath are not taken into account.

The relation above predicts a power dependence of the absorption
$\propto P^{-1/2}$ above some critical power $P_c$, although in
practice one may expect another high-power roll-off
once all the TLF are saturated and other loss mechanisms start to dominate.
The critical intensity is given by
$P_c=3\hbar^2 c_0\epsilon_0\epsilon_r/2 d^2 T_1 T_2$
where $T_1$ and $T_2$ are the relaxation and the dephasing times
of the TLF, respectively.

The variation in permittivity is dominated by resonant processes and can therefore be found by  applying
the Kramers-Kronig relations to
Eq.\ref{eq:absorption}
\begin{equation}
\frac{\epsilon(T)-\epsilon(T_0)}{\epsilon(T_0)}=-\frac{2nd^2
}{3\epsilon} \left[
\ln\left(\frac{T}{T_0}\right)-(g(T,\omega)-g(T_0,\omega)\right]
\label{eq:depsilon}
\end{equation}
where $g(T,\omega)=\textmd{Re} \Psi(\frac{1}{2}+\hbar \omega/2\pi
i k_B T)$, $T_0$ is a reference temperature and $\Psi$  the
complex digamma function significant only for $kT\lesssim\hbar
\omega/2$. From cavity pertubation theory we then have that
$\Delta f_0/f_0= -F/2 \Delta \epsilon/\epsilon$, $F$ being a
filling factor which depends on the geometry and the electric field
distribution. The substrate filling factor varies slightly between the three types of
resonator discussed here; but is over 0.9 in all cases. For samples on SiO$_2$/Si substrates the contribution of the SiO$_2$ layer to $F$ is about 0.2.
Eq.~\ref{eq:absorption}
predicts an upward turn  in the temperature dependence of the
centre frequency at mK temperatures, where the $g(T,\omega)$ term is
important. In the experiments by Gao \textit{et al} \cite{gao_upturn} Eq.~\ref{eq:depsilon} was used to fit the temperature dependence of the resonance frequency, but the low-temperature upturn was not clearly observed.

The aim of this experimental study is to verify whether the theory
developed for a spin solution in glass can accurately describe the
temperature and power dependence of the resonance frequency and
losses in superconducting resonators at mK temperatures.
\section{Experimental}
\begin{figure}[t]
\includegraphics[width=6cm]{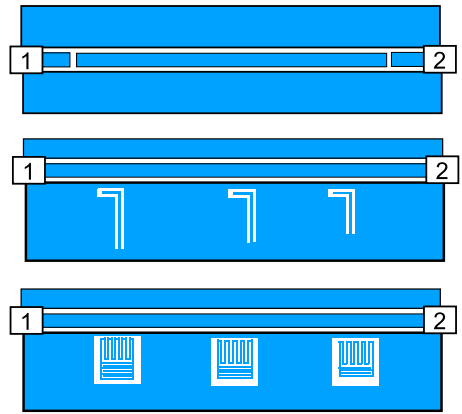}
\caption{\label{fig:LE} Illustration (\textit{not to scale}) showing the geometry
and measurement configuration (ports $2\rightarrow1$) for the three types of resonators used in this work. From top to bottom:
$\lambda/2$,$\lambda/4$ and Lumped element (LE) resonator. \label{fig:illustr}}
\end{figure}

We have fabricated a wide variety of $\lambda/2$, $\lambda/4$  and
lumped element (LE) niobium resonators (see fig.~\ref{fig:LE}) on R-cut sapphire and
thermally oxidized high-resistivity ($>1$ kOhm/cm$^2$) silicon
substrates with 400 nm of SiO$_2$. The simplest design is the in-line $\lambda/2$
resonator. Its advantages is that it can be made with loaded
quality factors of $>10^6$ and is easy to measure (due to the high
S/N ratio on resonance), there is however only one resonator
per chip and a calibrated measurement is needed in order to
quantify its unloaded quality factor (which is difficult at mK
temperatures). $\lambda/4$ resonators have the advantage that their parameters
can be extracted relatively easily (using numerical fitting) from  uncalibrated measurements since
the transmittance far from resonance can be used as a reference
level.  Several $\lambda/4$ resonators can be coupled to a common
 feedline, making it possible to
measure in a frequency-multiplexed configuration.  The lumped element resonator shares these advantages and
is much more compact (typical size is about 200x200 $\mu$m$^2$) than geometric resonators.
Our LE resonators fabricated in Nb on sapphire exhibit loaded
 Q close to  $10^5$ with an unloaded Q that is about
3 times higher; i.e the resonators are by design overcoupled.

The structures were patterned in 200 nm thick sputtered Nb films
using ion-beam etching and were designed to have their centre
frequencies in the c-band(4-8 GHz). Table \ref{tab:table} shows a
summary of the samples discussed in this paper. Other resonators
measured to-date showed similar behaviour.
\begin{table}[b]
\caption{\label{tab:table} Summary of the resonators used in these experiments presented in this paper}
\begin{ruledtabular}
\begin{tabular}{ccccc}
 &Type& Chip & Substrate &$f_0$ at 50 mK (MHz) \\
\hline
R1 & $\lambda/2$\footnotemark[1] & A & SiO$_2$/Si &6037  \\

R2 & $\lambda/4$ & B & SiO$_2$/Si & 5040  \\
R3 & $\lambda/4$ & B & SiO$_2$/Si  & 5795   \\
R4 & $\lambda/4$ & B & SiO$_2$/Si  & 8100   \\

R5 & LE & C & SiO$_2$/Si  &  6805 \\
R6 & LE & C & SiO$_2$/Si  &  7250  \\
R7 & LE & C & SiO$_2$/Si  &  7813  \\

R8 & LE & D & Sapphire &  6046 \\
R9 & LE & D & Sapphire &  6458  \\
R10& LE & D & Sapphire &  6959  \\

\end{tabular}
\end{ruledtabular}
\footnotetext[1]{ Single resonator on the chip,
measured in-line}
\end{table}
Each chip is 5x10 mm$^2$ in size and incorporates a central
50~$\Omega$ coplanar feed-line and one or more resonators. For the
measurement the chip is glued to a gold-plated alumina carrier
inside a superconducting box. All  measurements (with the
exception for the data above 1~K in fig. \ref{fig:lambda2}) were
carried out in a dilution refrigerator with heavily filtered microwave
lines. The transmitted signal was amplified using an InP HEMT
amplifier held at 1 K with a noise temperature of about 4~K,
before further amplification stages at room temperature.
\section{Results and discussion}
\begin{figure}[t]
\includegraphics[width=7.5cm]{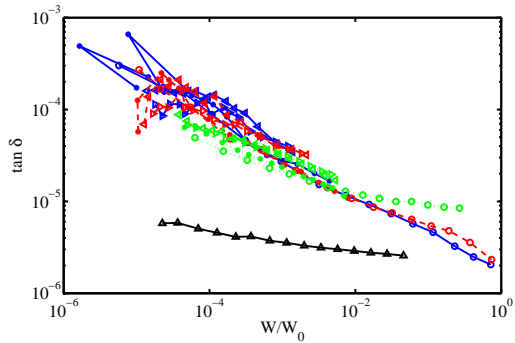}
\caption{\label{fig:tand} Loss tangent for 3 lumped element
resonators (R5-R7: green dotted, red dashed and solid blue line, respectively)  on SiO$_2$/Si at 4 different temperatures: 50mK($\circ$), 100mK ($*$), 250mK($\blacktriangleleft$) and 340mK($\blacktriangleright$).
The lowest curve ($\blacktriangle$) shows a similar measurement at 30mK for R8, a resonator on
sapphire. All curves were measured using the same \textit{applied} power.}
\end{figure}

Figure \ref{fig:tand} shows how the loss tangent
depends on the energy in the resonator on resonance $W=2Q_u
(1-S_{21})S_{21}P_{in}/\omega_0$ for three different lumped element
resonators on SiO$_2$/Si coupled to the same on-chip transmission line and
measured simultaneously.  Absolute power metrology at mK temperatures is very problematic so only relative measurements were possible; we estimate that the maximum power reaching the chip during the measurements is about -70~dBm, meaning $W_0$ is of the order of $10^{-16}$~ Ws.
Shown is also R8, a sapphire resonator of the same design. For the resonators fabricated on
Si$O_2$/Si the curves have a slope of approximately -0.4 in this range of applied powers. Eq.~\ref{eq:absorption} predicts
dielectric losses changing with a slope of -0.5 for powers above
$P_c$ and power-independent behavior below $P_c$, it is therefore
possible that we are in the intermediate regime. However, it is
also possible that losses in the superconductor not captured by
Eq.~\ref{eq:absorption} play a role. Indeed, the current distribution
in superconducting resonators is such that its \textit{local}
value can be significant even at our lowest
 powers leading to higher conductor losses as the power
is increased.
It is also interesting to note that the power entering the resonator depends
on Q \textit{and} Q -due to the TLF- in turn depends on the
power, $Q\propto P(Q)$. Hence, the system is non-linear and in principle
weakly hysteretic.

The main conclusion from  fig.~\ref{fig:tand} is that the loss
tangent is strongly power dependent for SiO$_2$/Si whereas a
weak dependence is observed for identical structures on
sapphire; in the latter case Q only changes by about a factor of two even when the
power is changed by over 40 dB.

\begin{figure}[b]
\includegraphics[width=7.5cm]{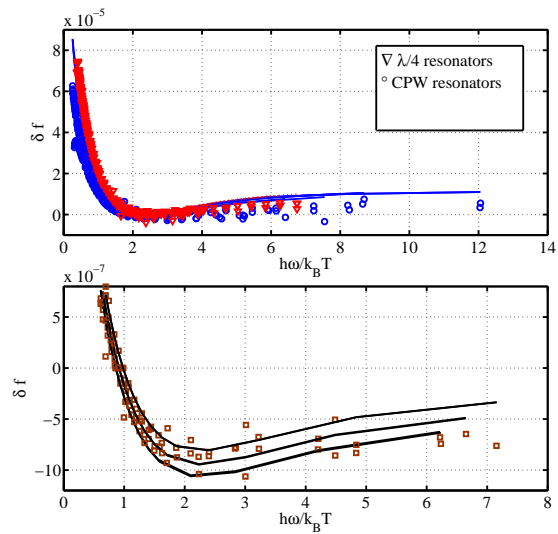}
\caption{\label{fig:fshift} Frequency shift as a function of the
normalized frequency at several different powers for LE resonators R2 to R7 on
SiO$_2$/Si (\textit{top});  and R8 to R10 sapphire(\textit{bottom}). The solid
lines show fits to theory. The $\log$ term dominates at small $\hbar \omega/k_BT$ whereas $g(t)$ becomes important in the opposite regime. Successive curves differ in power by 10 dB.}
\end{figure}

Experimentally, the temperature dependence of the resonance
frequency of superconducting resonators is quite complicated since
it is affected both by the properties of the superconducting film
(e.g. the kinetic inductance) and changes in the dielectric and it
is in general difficult to separate these two effects. However, at
temperatures $\ll T_c$, which is the case here, we can assume that the
thermal effects on the kinetic inductance are small.
Figure \ref{fig:fshift} shows the relative shift of the resonance
frequency as a function of  $\hbar
\omega/k_B T$ together with a fit to Eq.~\ref{eq:depsilon}.
All data from the resonators on a particular substrate can be fitted using a single  parameter;
we find $Fd^2n/3\epsilon=6\cdot10^{-5}$ and $2.8\cdot10^{-6}$
for the Si$O_2$/Si and sapphire respectively; a difference of about a
factor of twenty; consistent with the loss tangent measurements.
Note that the curves shown come from several resonators of two different designs (on separate chips), with each resonator
measured at several different power levels, but they can all be
fitted using a single parameter.
We note that the values for $Fd^2n/3\epsilon$ we obtain from samples fabricated on sapphire
are consistently lower than those reported by Gao \textit{et al}.

\begin{figure}[t]
\includegraphics[width=7.5cm]{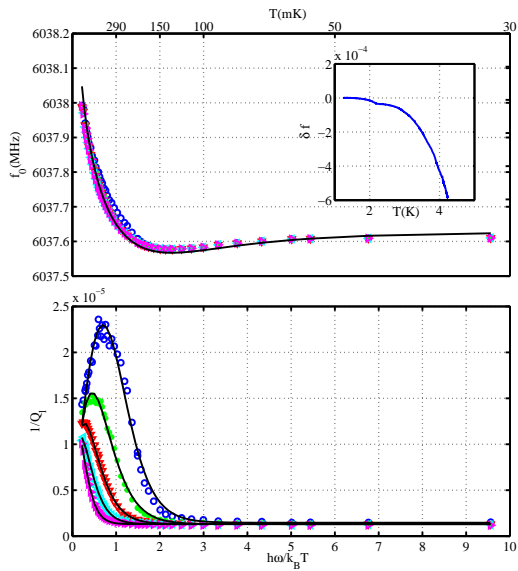}
\caption{\label{fig:lambda2} Centre frequency (\textit{Top}) and
$1/Q_l$ (\textit{Bottom}) at five different powers and as a
function of temperature for sample R1,  a $\lambda/2$ resonator
fabricated on SiO$_2$/Si. Each successive curve (top to bottom)
corresponds to an increase in power by 2 dB. Solid curves show fit
to theory. \textit{Inset:} $\lambda/2$ resonator measured above
1K.}
\end{figure}

In order to verify that the temperature dependence does not depend
on the geometry or our data analysis, we have in addition to
lumped-element and $\lambda/4$ (in shunt configuration) resonators
also measured undercoupled in-line $\lambda/2$ coplanar
resonators. Fig.~\ref{fig:lambda2} shows the measured temperature
dependence of 1/$Q_l$ -$Q_l$ being the loaded quality factor-  and
$f_0$. Again we can fit all the $f_0$ curves using a single
parameter, although in this case the fitting parameter
$Fd^2n/3\epsilon$ is $4.95\cdot10^{-5}$, slightly lower than for
the other samples. Some of this difference is likely to be due to
somewhat different filling factors.

At temperatures above 1K the losses are
dominated by the conductor losses described by the Mattis-Bardeen
theory but at low temperatures and low microwave power conductor
losses are no longer important; resonant absorption dominates
and the losses increase as the temperature is reduced as
captured by Eq.\ref{eq:absorption}. High power on the other hand
saturates the TLS thus suppressing the resonant absorption.
Fig.~\ref{fig:lambda2} (\textit{bottom}) illustrates the regime of
decreasing losses at temperatures below about 300 mK, which can
only be observed in a very narrow range of microwave power. This
regime is obviously not described by resonant absorption
Eq.~\ref{eq:absorption}. The likely explanation is that the
relative significance of \textit{relaxation} absorption
\cite{Galperin} processes increases for temperatures below
$\approx \hbar \omega /k_B$ and intermediate power levels, just
enough to suppress the resonant absorption. As the power is
increased, the crossover to the relaxation losses is pushed to
higher temperatures.

Whilst the presence of TLF is  known to affect
resonators it is nevertheless surprising how well our data can be fitted
using a single parameter independent of both
the frequency (over a range of several GHz) and the location on the
chip. While our data do not give us a reliable way to
identify the nature of these TLF, one can with some certainty
conclude that their origin must somehow be intrinsic to the
materials used. The bath of TLF appears to have a very
wide distribution of both the tunnel splittings  and the symmetry parameter of the
(effective) double-well. The large $nd^2$ factor in SiO$_2$/Si
-consistent with previous measurements\cite{Oconnell}- comes as no surprise given the large density of
TLF one would expect in an amorphous oxide material. The
$nd^2$ factor in sapphire is significantly smaller, but still
measurable. This is in itself quite interesting as it is well
known that the only two-levels systems seen in very high-quality
dielectric sapphire resonators are known to be due to -relatively
narrow- electron spin resonances on incidental paramagnetic ions,
such as Fe or Cr; if these were the orgin of the TLF in our samples one would not
expect the data to agree with the theory used here.
This suggests that the TLF seen here are located
in the interface between the film and the substrate, as opposed to
the bulk material.
\begin{acknowledgments}
We wish to acknowledge John Gallop, Yuri Galperin, Gregoire
Ithier, Olga Kazakova, Phil Mausskopf, Phil Meeson, Mark Oxborrow,
Giovanna Tancredi and members of NTT Basic Research Laboratories
for helpful discussions and comments. This work was supported by
EPSRC and the the National Measurement Programme of the Department for Industries, Universities and Skills (DIUS).
\end{acknowledgments}
\bibliography{tempdep_CPR}
\end{document}